\documentstyle[amssymb,preprint,aps]{revtex}

\begin{document}
\title{The cosmological origin of Higgs particles}
\author{Liao Liu \ \ \ Shouyong Pei}
\address{Department of Physics, Beijing Normal University, Beijing 100875,\\
China}
\maketitle

\begin{abstract}
A proposal of the cosmological origin of Higgs particles is given. We show,
that the Higgs field could be created from the vacuum quantum conformal
fluctuation of Anti-de Sitter space-time, the spontaneous breaking of vacuum
symmetry, and the mass of Higgs particle are related to the cosmological
constant of our universe,especially the theoretical estimated mass m$_{H}$
of Higgs particles is m$_{H}=\sqrt{-2\mu ^{2}}$ =$\sqrt{|\Lambda /\pi }$.
\end{abstract}

\pacs{PACS number(s): 1100N, 9880}

we know, Higgs field is so important in modern field theory, the spontaneous
breaking of vacuum symmetry of it, and the associated Higgs mechanism are
essential in the establishment of a physical reasonable modern gauge field
theory. However now after 37 years of Higgs' et al pioneer work\cite{ref1},
Higgs field is still some thing put in 'by hand' in field theory, no one
knows the origin of it, no one knows how to give a theoretical estimation of
it's mass, etc. So a possible proposal of the origin of it might have
fundamental significance in modern physics.

We shall show in this short report that Higgs field could be created from
vacuum quantum conformal fluctuation (VQCF) of anti-de Sitter space-time.

Let the VQCF of certain background space-time $g_{ab}$, devoid of matter is $%
\overline{g}_{ab}$, i.e. \ 
\begin{equation}
\overline{g}_{ab}=\left\langle 0\left| \Omega ^{2}\right| 0\right\rangle
g_{ab}\equiv \phi ^{2}g_{ab}
\end{equation}%
where $\Omega $ is an operator, $\phi ^{2}\equiv \left\langle 0\left| \Omega
^{2}\right| 0\right\rangle $ is the vacuum expectation value of $\Omega ^{2}$
in certain vacuum quantum state 
\mbox{$\vert$}%
0$\rangle $ of space-time $g_{ab}$. Then the Einstein-Hilbert action $%
\overline{I}\left( \overline{g}_{ab}\right) $ of VQCF of the
Einstein-Hilbert action $I\left( g_{ab}\right) $ will be 
\[
\bar{I}\left( \bar{g}_{ab}\right) =\frac{1}{16\pi G}\int \left( \bar{R}%
-2\Lambda \right) \left( -\bar{g}\right) ^{\frac{1}{2}}dx^{4}
\]%
\begin{equation}
=\overline{I}\left( \varphi ,g_{ab}\right) =\frac{1}{16\pi G}\int \left(
-6\phi _{;a}\phi _{;b}g^{ab}+R\phi ^{2}-2\Lambda \phi ^{4}\right) \left(
-g\right) ^{\frac{1}{2}}dx^{4}
\end{equation}%
in signature $(-2)$ \cite{ref2}\cite{ref3}. No doubt, this action is a
physical reasonable Einstein-Hilbert action of the conformal equivalent
space-time $\overline{g}_{ab}$, but if one tries to look upon Eq. (2) as the
action of $\lambda \varphi ^{4}$ scalar field in certain background
space-time $g_{ab}$, we shall find there is a wrong minus sign in the
kinetic term as is evident by comparing Eq. (2) with that of $\lambda
\varphi ^{4}$ scalar Higgs field 
\[
I\left( \varphi ,g_{ab}\right) =\int \left( \frac{1}{2}\varphi _{;a}\varphi
_{;b}g^{ab}-\frac{\mu ^{2}}{2}\varphi ^{2}-\frac{\lambda }{4}\varphi
^{4}\right) \left( -g\right) ^{\frac{1}{2}}dx^{4}
\]%
where $\mu ^{2}<0,\quad \lambda >0$ and the signature is (-2). Such
situation implies, though Eq. (2) is really a physical reasonable gravity
theory for the conformal equivalent space-time $\overline{g}_{ab}$, it isn't
a physical reasonable action of $\lambda \varphi ^{4}$ scalar field in
background space-time $g_{ab}$.

Now we shall show there exists a modified action other than Eq. (2) that can
give the same gravitational field equation as Eq. (2) does, moreover, it can
simultaneously give a physical reasonable meaning to the\ action of the $%
\lambda \varphi ^{4}$ scalar field in background space-time $g_{ab}$.

The modified action satisfied the above two demands is simply the action
obtained by changing the sign of Eq. (2) to minus, i.e. 
\begin{eqnarray}
L\left( \phi \right)  &=&\frac{1}{16\pi G}\left( 6\phi _{;a}\phi
_{;b}g^{ab}-R\phi ^{2}+2\Lambda \phi ^{4}\right)   \nonumber \\
&=&\frac{1}{2}\left( \frac{3}{4\pi G}\right) \phi _{;a}\phi _{;b}g^{ab}-%
\frac{R}{16\pi G}\phi ^{2}+\frac{\Lambda }{8\pi G}\phi ^{4}
\end{eqnarray}%
Put%
\begin{equation}
\frac{R}{16\pi G}=\frac{1}{2}\mu ^{2}\quad or\quad \mu ^{2}=\frac{R}{8\pi G}
\end{equation}%
\begin{equation}
\frac{\Lambda }{8\pi G}=-\frac{\lambda }{4}\quad or\quad \lambda =-\frac{%
\Lambda }{2\pi G}
\end{equation}%
then Eq. (3) becomes%
\begin{eqnarray}
L\left( \phi \right)  &=&\frac{1}{2}\left( \frac{3}{4\pi G}\right) \phi
_{;a}\phi _{;b}g^{ab}-\frac{1}{2}\mu ^{2}\phi ^{2}-\frac{\lambda }{4}\phi
^{4}=\frac{1}{2}\left( \frac{3}{4\pi G}\right) \phi _{;a}\phi
_{;b}g^{ab}-U\left( \phi \right)  \\
U\left( \phi \right)  &=&\frac{1}{2}\mu ^{2}\phi ^{2}+\frac{\lambda }{4}\phi
^{4}
\end{eqnarray}%
Here $\mu ^{2}<0,\quad \lambda >0$ if $\Lambda <0,\quad R=4\Lambda <0$.
Clearly Eq.(6) is exactly the Lagrangian of $\lambda \phi ^{4}$\ Higgs
field, except there is an extra factor $\left( \frac{3}{4\pi G}\right) $ in
the kinetic term.

Thus at last we find that Higgs field is just the vacuum\ quantum conformal
fluctuation of the background anti-de Sitter space-time, the imaginary mass $%
\mu $ and the self-interaction constant $\lambda $ are determined solely by
the gravitational constant and the cosmological constant of the background
anti-de Sitter space-time.

Especially, it is interesting to note that the spontaneous breaking of
vacuum symmetry is just a result of vacuum quantum conformal fluctuation of
anti-de Sitter space-time.

Now, a comment should be given, as our universe is certainly not an anti-de
Sitter (AdS) universe, so the VQCF of it and all the above associated
conclusion may be nonsense. However we would like to point out, according to
the description of the dynamics of superbrane theory, superbrane may be
propagated in the background of Anti-de Sitter space-time\cite{ref4}.

Also, we will show in the Appendix that anti-de Sitter space-time of $\left(
-\Lambda \right) $ might be created through quantum process or phase
transition from de Sitter space-time of $\Lambda $.

At last, as in conventional gauge field theory, one can easily give the
following modified masses $m_{H}$, $m_{l}$, and $m_{g}$ of Higgs particle,
lepton and gauge particles respectively in c=$\hslash $=G=1 unit%
\begin{equation}
m_{H}=\sqrt{-2\mu ^{2}}=\sqrt{-\frac{R}{4\pi G}}=\sqrt{\frac{\left| \Lambda
\right| }{\pi }}\qquad \qquad
\end{equation}%
\begin{equation}
m_{l}=G_{l}\sigma =G_{l}\quad \quad
\end{equation}%
\begin{equation}
m_{g}=\sqrt{\frac{3}{4\pi G}g^{2}\sigma ^{2}}=\sqrt{\frac{3}{4\pi }}g\cdot
\sigma
\end{equation}%
note, $\sigma =\sqrt{-\frac{\mu ^{2}}{\lambda }}=\sqrt{\frac{R}{4\Lambda }}%
=1 $ from vacuum Einstein equation with $\Lambda $-term. $G_{l}$, g are the
Yukawa coupling constant and gauge coupling constant respectively. Eq. (9)
and Eq. (10) show that the masses of leptons and gauges particles are of the
same order as that in conventional gauge field theory, but the mass of Higgs
particle is related to the cosmological constant of our universe from Eq.
(8), so if we know $\Lambda ,$ we can determine m$_{H}$ and vice versa.

As was shown by S. Perlmutter et al in 1999, \cite{ref5}, a best-fit flat
cosmology to the observed Hubble constant $H_{0}$ and cosmological constant $%
\Lambda $ from the red-shift of $I_{a}$-type supernova, that is%
\begin{eqnarray}
\Omega _{M}+\Omega _{\Lambda } &=&1  \nonumber \\
(\Omega _{M,}\text{ }\Omega _{\Lambda )} &=&\left( 0.28,0.72\right) 
\end{eqnarray}%
where $\Omega _{M,\Lambda }=\frac{\rho _{M,\Lambda }}{\rho _{c}},\rho _{c}=%
\frac{3H_{0}^{2}}{8\pi G}$is the critical mass density of our universe. It
is straightforward to show from (11), $\Omega _{\Lambda }=\frac{\Lambda }{%
3H_{0}^{2}}$ and $H_{0}\simeq \left( 50-80\right) km\cdot s^{-1}\left(
Mpc\right) ^{-1}$ to get the recent value of  $\Lambda =\left(
6.66-17\right) \times 10^{-58}cm^{-2}.$So the recent value $m_{H}\quad
\simeq \left( 0.39-1\right) \times 10^{-33}ev$ from Eq. (8) which is about 10%
$^{-27}$ times the mass of electron! That is too small as expected by
particle physicists.

However, we should point out, until now, no one can give an universal agreed
lower limit of m$_{H}$. S. Weinberg in 1996 concluded that two groups in
1995 could only give an upper limit of m$_{H}\prec $ 225Gev\cite{ref6}. The
LEP group seems to obtain in 1996-2000 a lower limit m$_{H}\succ $108Gev or
114Gev$\prec $m$_{H}\prec $ 118Gev. But the works of four groups following
them in LEP got nothing to confirm the above results, as we know this leads
to the final close of LEP\cite{ref7}. As we know there is an inflationary
phase in the very early evolution of our universe, when the cosmological
constant $\Lambda $ or the vacuum energy density $\frac{\Lambda C^{4}}{8\pi G%
}$ may be very large as compared to the nowadays' value $\Lambda _{0}\simeq $
$10^{-58}cm^{-2}$,e.g. Linde estimated $\Lambda $ may be as large as 3$%
\times 10^{40}cm^{-2}$\cite{ref8}. the corresponding Higgs' mass from Eq.
(8) will be $\simeq $10$^{7}Gev$. If $\Lambda \simeq $10$^{34}$ $cm^{-2},$%
the corresponding Higgs' mass from Eq. (8) will be just the above mentioned
lower limit 108$Gev$, as the cosmological constant $\Lambda $ is really a
slowly decreasing quantity in the inflationary era, so it seems that the
Higgs particles with masses in Gev range may be the relics of the VQCF of
Anti-de Sitter space in inflationary phase of our very early universe. This
is a very interesting result.

\appendix Appendix

First, let the Euclidean section of de Sitter instanton reads%
\begin{equation}
ds^{2}=\frac{1}{H^{2}}\left( d\theta ^{2}+\sin ^{2}\theta d\Omega
_{3}^{2}\right)  \eqnum{A-1}  \label{test}
\end{equation}%
where%
\[
H^{2}=\frac{\Lambda }{3},\quad \Lambda >0 
\]%
$\Omega _{3}^{2}$ is the unit 3-sphere.

We perform first a analytical continuation at the equatorial space-like
hypersurface of $S^{4}$ or Eq. (A-1) where the second fundamental form
vanishes. 
\begin{equation}
\theta \longrightarrow \frac{\pi }{2}-i\chi  \eqnum{A-2}
\end{equation}%
we obtain then a Lorentzian inflationary closed universe%
\begin{equation}
ds^{2}=\frac{1}{H^{2}}\left( -d\chi ^{2}+\cosh ^{2}\chi d\Omega
_{3}^{2}\right)  \eqnum{A-3}
\end{equation}%
or an Euclidean closed space%
\begin{equation}
ds^{2}=\frac{1}{H^{2}}\left( d\overline{\chi }^{2}+\cos ^{2}\overline{\chi }%
d\Omega _{3}^{2}\right)  \eqnum{A-4}
\end{equation}

\bigskip After putting 
\[
\chi \longrightarrow -i\bar{\chi} 
\]

The second step analytical continuation is performed at $\overline{\chi }=%
\frac{\pi }{2}$. or%
\[
\overline{\chi }\longrightarrow \frac{\pi }{2}-i\Phi 
\]%
we then obtain from Eq. (11) an AdS space\cite{ref9}%
\begin{equation}
ds^{2}=\frac{1}{H^{\prime 2}}\left( d\Phi ^{2}+\sinh ^{2}\Phi d\Omega
_{3}^{2}\right)   \eqnum{A-5}
\end{equation}%
where%
\[
H^{\prime 2}=-H^{2}=-\frac{\Lambda }{3}<0
\]%
so if we are now live in an inflating de Sitter bubble of $\Lambda >0$, AdS
spaces of $\left( -\Lambda \right) <0$ may be created through quantum
process or phase transition from de Sitter space\cite{ref9}, and then Higgs
fields created from the VQCF of AdS space.\cite{ref5}

\bigskip we should note, the boundary $\overline{\chi }=\frac{\pi }{2}$ or $%
\Phi =0$ between space (A-4) and space (A-5) is a null surface, which is
really a Cauchy horizon separating the causal de Sitter region (A-4) from
the acausal AdS region (A-5), here the junction problem exists on this
surface, as was shown in references\cite{ref9}, no junction trouble exists
on time-like and space-like boundary, as the null surface may be looked upon
as the limiting case of non-null surface, we may conjecture no junction
trouble exists also in the above second step continuation, though a rigorous
proof is required.

\bigskip

This paper is supported by the Natural Science Foundation of China under
Grant No. 19473003. The authors wish to thank Dr. B. B. Wang, and Dr. Y. G.
Ma for helpful discussions.

E-mail address: LiuLiao1928@yahoo.com.cn


\begin{references}
\bibitem{ref1} P. W. Higgs, Phys. Lett. {\bf 12}, 132 (1964), Phys. Rev.
Lett. {\bf 13}, 508 (1964)

\bibitem{ref2} T. Padmanabhan, Phys. Rev. {\bf 28}D, 745 (1983).

\bibitem{ref3} S. W. Hawking, in General relativity, An Einstein century
survey edited by S. W. Hawking \ and W. Israel, Cambridge university Press
1979

\bibitem{ref4} D. Sorokin, Physics Report, 329 No 1/2 (2000), E. Witten,
J.High Energy Phys. 7 006(1998)

\bibitem{ref5} S. Perlmutter, et al, APJ {\bf 517, \ }565 (1999)

\bibitem{ref6} S. weinberg, Quantum Theory of field,(1996) Vol II, p316,
Cambridge University Press

\bibitem{ref7} CERN COURIER, 40(10)5(2000)

\bibitem{ref8} A. D. Linde, Phys. Lett. 99B 391(1981)

\bibitem{ref9} L. X. Li, Phys Rev 59D, 084016(1999)
\end{references}
\end{document}